\documentclass[12pt]{article}

\usepackage{latexsym}
\usepackage{amsmath}
\usepackage{amsfonts}
\usepackage{amssymb}
\usepackage{psfrag}
\usepackage{graphicx}
\usepackage{longtable}
\usepackage{caption}
\usepackage{hyperref}
\usepackage{color}
\usepackage{authblk}

\def \ve{\mbox{\boldmath $e$ \unboldmath}\!\!}

\def \vW{\mbox{\boldmath $W$ \unboldmath}\!\!}

\def \vSigma{\mbox{\boldmath $\Sigma$ \unboldmath}\!\!}

\def \vLambda{\mbox{\boldmath $\Lambda$ \unboldmath}\!\!}

\def \vbeta{\mbox{\boldmath $\beta$ \unboldmath}\!\!}
\def \veta{\mbox{\boldmath $\eta$ \unboldmath}\!\!}

\def \vgamma{\mbox{\boldmath $\gamma$ \unboldmath}\!\!}
\def \vepsilon{\mbox{\boldmath $\epsilon$ \unboldmath}\!\!}

\def \vb{\mbox{\boldmath $b$ \unboldmath}\!\!}

\def \vxi{\mbox{\boldmath $\xi$ \unboldmath}\!\!}

\def \vx{\mbox{\boldmath $x$ \unboldmath}\!\!}

\DeclareMathOperator*{\yl}{\mathbf{y}_{\ell}}

\DeclareMathOperator*{\argmin}{arg\;min}  % keep
\DeclareMathOperator*{\mlikelihood}{\sum_{\ell = 1}^n || \mathbf{W}^T\mathbf{x}_{\ell} - \mathbf{y}_{\ell}||_2^2}

\makeatletter     %%% keep
\newcommand{\distas}[1]{\mathbin{\overset{#1}{\kern\z@\sim}}}%
\newcommand{\distras}[1]{%
  \savebox{\mybox}{\hbox{\kern3pt$\scriptstyle#1$\kern3pt}}%
  \savebox{\mysim}{\hbox{$\sim$}}%
  \mathbin{\overset{#1}{\kern\z@\resizebox{\wd\mybox}{\ht\mysim}{$\sim$}}}%
}
\makeatother

\begin{document}
%\firstpage{1}
\author[$^{1}$]{Farouk S. Nathoo}
\author[$^{2}$]{Linglong Kong}
\author[$^{3}$]{Hongtu Zhu}

\affil[$^{1}$]{Department of Mathematics and Statistics, University of Victoria}
\affil[$^{2}$]{Department of Mathematical and Statistical Sciences, University of Alberta}
\affil[$^{3}$]{Department of Biostatistics, The University of Texas MD Anderson Cancer Center}

%$^{1}$Department of Mathematics and Statistics, University of Victoria\\
%$^{2}$Department of Mathematical and Statistical Sciences, University of Alberta\\
%$^{3}$Department of Biostatistics, The University of Texas MD Anderson Cancer Center\\
%$^{*}$nathoo@uvic.ca\\

\title{A Review of Statistical Methods in Imaging Genetics}
%\author[Sample \textit{et~al}]{Keelin Greenlaw$^{1}$, Elena Szefer\,$^{2}$, Jinko Graham\,$^{2}$, Mary Lesperance$^{1}$, and Farouk S. Nathoo\,$^{1,}$\footnote{to whom correspondence should be addressed}; For the Alzheimer's Disease Neuroimaging Initiative}
%\address{$^{1}$Mathematics and Statistics, University of Victoria, Victoria, British Columbia, PO BOX 1700 STN CSC, Canada.\\
%$^{2}$Statistics and Actuarial Science, Simon Fraser University, Burnaby, British Columbia, V5A 1S6, Canada.}

\maketitle

\begin{abstract}

With the rapid growth of modern technology, many large-scale biomedical  studies    have been/are being/will be conducted to collect  massive datasets with large volumes of   multi-modality imaging, genetic, neurocognitive, and clinical information from increasingly large cohorts. Simultaneously extracting and integrating rich and diverse heterogeneous information in neuroimaging and/or genomics from these big   datasets  could transform our understanding of how genetic variants impact brain structure and function, cognitive function, and brain-related disease risk across the lifespan. 
Such understanding  is critical 
for    diagnosis, prevention, and treatment of numerous complex brain-related disorders (e.g., schizophrenia and Alzheimer's disease).     
However,  
  the development of analytical methods for the joint analysis of both  high-dimensional imaging phenotypes  and high-dimensional genetic data, referred to as 
  big data squared (BD$^2$),    
presents  
major computational and theoretical challenges for    existing analytical methods.   
Besides the high-dimensional nature of BD$^2$,  
 various neuroimaging measures often exhibit strong spatial smoothness and dependence and  genetic markers may have a natural dependence structure arising from linkage disequilibrium. 
 We review some recent developments of various statistical techniques for  the joint analysis of BD$^2$,  including massive univariate and voxel-wise approaches, reduced rank regression, mixture models,  and group sparse multi-task regression. By doing so,  we hope that this review may encourage others in the statistical community to enter into this new and exciting field of research.

\end{abstract}

\section{INTRODUCTION}

%{\color{red}
Despite the numerous successes of genome-wide association studies (GWAS), 
it has been difficult to unravel the genetic basis of many  complex neurological diseases since each genetic variant may only contribute  in a small way to disease risk and such a  genetic basis can be very heterogeneous (Cannon and Keller, 2006; Marenco and Radulescu, 2010; Peper, et al. 2007).
%\cite{Cannon2006,Turner2006,Scharinger2010,Marenco2010,Rasch2010,Domschke2010,
%Bigos2010,Pezawas2010,Durston2010,Inkster2010,Casey2010,Paus2010,Meyer-Lindenberg2009,Glahn2007,
%Peper2007}.    
%Most psychiatric disorders begin to appear in childhood and adolescence and then begin to ebb. Their origins are in childhood and even in prenatal brain development (Satterthwaite, et al., 2014 ). 
%\cite{bale2010early,satterthwaite2014neuroimaging,paus2008many,kessler2005lifetime}. 
%Therefore, major psychiatric disorders often are regarded as the end result of abnormal trajectories of childhood brain development. 
The additive and interactive effects of perhaps hundreds of risk genes and multiple environmental risk factors, each with small individual effects, may contribute to the abnormal developmental trajectories that underlie neurological and psychiatric disorders such as Alzheimer's Disease. Identifying such risk genes and environmental risk factors could transform our understanding of the origins of these conditions and inspire new approaches for urgently needed preventions, diagnoses, and treatments. Once such an identification has been accomplished, lifestyle and medical interventions can be applied to make a potential difference in the outcome. 

A promising approach to  understanding the genetic basis of neurological disorders is through studies that integrate multi-scale data from genetic/genomic, multimodal brain imaging,  and  environmental risk factors (Hibar, et al., 2011; Thompson, et al., 2013; Hibar, et al., 2015), so called imaging genetics studies. 
%\cite{Hibar2011,Chen2012,Ge2012,Thompson2013,Medlan2014,ge2015massively,ge2015kernel,Hibar2015,LinWang2014}.      

To promote such studies, the Brain Imaging Clinical Research Program at the National Institute of Mental Health (NIMH) has called for the establishment of relationships between genetic variations and imaging and cognitive findings and phenotypes in adult mental disorders.  To this end, a number of large-scale publicly available imaging genetic  databases have been established,     including  the Human Connectome project (HCP) study,
  the UK biobank  (UKbb) study,  
 the Pediatric Imaging, Neurocognition, and Genetics (PING) study,    
 the Philadelphia Neurodevelopmental Cohort (PNC),  and  the Alzheimer's Disease Neuroimaging Initiative (ADNI) study,  among many others. The ADNI database in particular has been used extensively by statisticians working on the development of methods for the joint analysis of neuroimaging and genetic data, and this database serves as a good starting point for new researchers in the area.

In such studies, the data available for each subject may include multiple 
MRI images, such as structural MRI, diffusion tensor imaging (DTI), and functional MRI,   cognitive assessments, and   genomic data (e.g.,   SNP array and copy number variations (CNVs)).   
 Jointly analyzing   imaging genetics   with clinical variables, however,  raises serious challenges as  existing statistical methods are rendered  infeasible for 
 efficiently analyzing  large-scale imaging genetic data sets  with many subjects. 
 These challenges arise from a setting where the data involve  high-dimensional  imaging data $\times$ high-dimensional genetic data --  
so-called Big Data squared (BD$^2$), complex correlation and spatial structures within both imaging and genetic data, and a potentially large number of subjects.

 For many brain-related diseases,   
since changes in brain structure and function are very subtle,  it is common to normalize 
multi-modal neuroimaging data to a common template (Xu, et al., 2003; Miller and Younes, 2001).
 %\cite{Xu2003,miller2001group}.   
 After normalization, 
  various  imaging phenotypes are commonly calculated from structural and functional imaging  data  (Friston, 2009; Zhu, et al., 2007). 
%\cite{Friston2009, Friston2007,   Huettel2004,Lindquist2008b, Nichols2003,Thompson2002,Grenander1998,  Chung2005, Grenander2007,Chung2008,Styner2005,Styner2004, Basser1994b,  Zhu2007b}.   
These normalized neuroimaging phenotypes   are  functional data measured at a very large number ($10^4-10^7$) of grid points along space and/or time
and network data measured among a large number ($10^4-10^6$) of  region of interest (ROI) pairs  (Hibar, et al., 2011; Thompson, et al., 2013; Hibar, et al., 2015; Ge, et al., 2015a, 2015b). 
%\cite{Hibar2011,Chen2012,Ge2012,Thompson2013,Medlan2014,ge2015massively,ge2015kernel,Hibar2015}.  
See Figure 1  for a graphical depiction of potential Imaging Phenotypes (IPs).
%}

%Brain imaging and genetic data are being increasingly collected and combined in order to study and diagnose inherited diseases including common mental disorders and neurodegenerative disorders (e.g., Alzheimer's Disease). It is possible to carry out a joint analysis of the brain imaging and genetic data by treating the brain imaging data as phenotypes for complex disorders. In such studies the primary interest lies with examining associations between genetic variations and quantitative traits derived through neuroimaging, the latter of which yields an image(s)  summarizing the structure or function of the brain. 

%This combination of potentially high-dimensional data types, where the response is an image and the covariates comprise a potentially massive number of genetic markers, is receiving increased attention in neuroscience, and the field of imaging genetics is catching up with the dramatic increase in the number of genome-wide association studies (GWAS). 

%Compared to studies examining more traditional phenotypes such as case-control status or cognitive test scores, the endophenotypes derived through neuroimaging are in some cases considered closer to the underlying etiology of the disease being studied, and this may lead to more powerful identification of the important genetic variations. This is particularly important for GWAS where large sample sizes are generally required in order to obtain reproducible results due to the very small effects conferred by individual loci (Munafo and Flint, 2011).

The earliest methods developed for imaging genetics data analysis are either based on significant reductions to both data types, for example, restricting the analysis to a specific candidate ROI in the brain and/or a specific candidate genetic marker. This type of univariate analysis can be extended to handle full brain-wide genome-wide data based on the application of a massive number of pairwise univariate analyses, each based on a standard linear regression relating a given voxel/region to a given SNP. In this case the multiple testing problem can be on a very large scale and the resulting corrections very stringent, given the large number of tests involved. For example, in a full brain-wide genome-wide study involving $10^6$ known genetic variants and $10^6$ locations in the brain, this type of analysis will require $10^{12}$ univariate analyses. Furthermore, the resulting p-values are not independent because of spatial correlation in the imaging data. 

Stein et al. (2010) are the first to consider such an analysis and these authors examined $448,293$ SNPs in each of $31,622$ voxels of the entire brain across 740 elderly subjects with 300 computational-cluster nodes used to carry out the required computations in parallel. Hibar et al. (2011) consider a similar analysis but reduce the number of tests by conducting the analysis at the gene rather than SNP level. In this case principal component analysis is used to summarize the SNP data for each gene, and the resulting `eigenSNPs' are used in the massive univariate analysis. 

As an alternative to the massive univariate approach, a voxel-wise approach continues to fit regression models separately at each location in the brain, but considers a set of genetic markers simultaneously rather than just a single genetic marker. Ge et al. (2011) develop such an analysis and examine a dataset that is similar to that considered in Stein et al. (2010), but a key difference is the use of a multi-locus model based on least squares kernel  machines (Liu et al., 2007), which is used to combine the effect of multiple genetic variants and model their interaction. In addition, the spatial information in the images is accounted for through the use of random field theory as an inferential tool (Worsley, 2002). This approach is extended in Ge et al. (2015) to allow for potential interactions between genetic variables and non-genetic variables such as disease-risk factors, environmental exposures, and epigenetic markers. 

An alternative fast  voxel-wise genome-wide association analysis (FVGWAS) approach is that developed by Huang et al. (2015) where the authors focus on reducing the computational burden required for a full brain-wide gene-wide study. This objective is implemented in part by incorporating a global sure independence screening procedure along with inference based on the wild bootstrap. The resulting approach can implement a brain-wide genome-wide analysis in a relatively small amount of time utilizing only a single CPU. 

One drawback of the massive univariate and voxel-wise approaches is that the relationship between the different neuroimaging phenotypes (e.g. at different regions of the brain) is not explicitly modelled, and therefore, potential efficiency gains arising from the borrowing of information across brain regions are not realized. An alternative approach is to base the analysis on a single large model, a multivariate high-dimensional regression model that is fit to the entire dataset. In this framework the scale of the data must necessarily be reduced, and it is common to summarize the entire image using a relatively moderate number of brain summary measures across some key ROIs. As an example, Table 1 describes a phenotype of dimension $56$ that can be derived from an MRI image, and these data are considered in our example application. 

One such regression model is the group-sparse multitask regression model proposed by Wang et al. (2012) where estimation of the regression coefficients in a multivariate linear model is based on penalized least squares. The penalty is chosen to induce a particular form of structured sparsity in the solutions based on two nested forms of group sparsity. The first is at the SNP level (grouping the regression coefficients of a given SNP across all phenotypes) and the second is at the gene level, which groups all SNPs within a given gene. More recently, Greenlaw et al. (2017) have extended this approach to the Bayesian setting which allows for inference and uncertainty quantification for the regression coefficients of the selected genetic markers. 

An alternative multivariate approach is based on approximating the high-dimensional regression coefficient matrix with a low rank matrix. Such an approach has been developed by Vounou et al. (2010), who develop a sparse reduced-rank regression (SRRR) model which is applied to an imaging genetics study involving $111$ anatomical ROIs and $437,577$ SNPs. Using simulation studies Vounou et al. (2010) show that their SRRR model has higher power to detect deleterious genetic variants compared with the massive univariate approach. Along similar lines, Zhu et al. (2014) also develop a low rank regression model with inference conducted in the Bayesian framework and they apply their approach to an imaging genetics study involving $93$ ROIs and $1,071$ SNPs. Also in the Bayesian framework, Stingo et al. (2013) develop a hierarchical mixture model for relating brain connectivity to genetic information for studies involving functional magnetic resonance imaging (fMRI) data. The mixture components of the proposed model are used to classify the study subjects into subgroups, and the allocation of subjects to these mixture components is linked to genetic markers with regression parameters assigned spike-and-slab priors. The proposed model is used to examine the relationship between functional brain connectivity based on fMRI data and genetic variation.

%{\color{red} 
Huang et al. (2017) developed a 
 functional genome-wide   association    analysis (FGWAS) framework to efficiently carry  out whole-genome analyses of functional phenotypes.
     Compared with FVGWAS, FGWAS explicitly
   models the functional features of   functional phenotypes through the integration of
    smooth coefficient functions and functional principal component analysis. Statistically, compared with existing methods for  genome-wide association studies (GWAS),   FGWAS can  substantially boost  the detection power for discovering important genetic variants  influencing    brain structure and function. 
%    }

In more recent work, researchers have turned their attention to longitudinal imaging genetics studies where study subjects are followed over time with neuroimaging data collected over a sequence of time points during a follow-up period. With longitudinal MRI data, changes in the structure of the brain over time can be characterized, for example, by examining rates of brain deterioration, and these estimated rates of change can be related to genetic markers. Szefer et al. (2017) examine the presence of linear association between minor allele counts of $75, 845$ SNPs in the Alzgene linkage regions and estimated rates of change of structural MRI measurements for 56 brain regions. The authors develop a bootstrap-enhanced sparse canonical correlation analysis to create refined lists of SNPs associated with rates of structural change over time. 

Lu et al. (2017) develop a Bayesian approach to perform longitudinal analysis of multivariate neuroimaging phenotypes and candidate genetic markers obtained from longitudinal studies. A low rank longitudinal regression model is specified where penalized splines are incorporated to characterize an overall time effect, and a sparse factor analysis model coupled with random effects is proposed to account for spatiotemporal correlations of longitudinal phenotypes. A useful feature of the proposed methodology is the allowance for interactions between genetic main effects and time.

In the remainder of the paper, Sections 2 - 4 discuss in more detail some of the methods mentioned above, with emphasis placed on our own work. Section 5 presents an example application where data from the ADNI-1 database are used to examine the association between the 56 neuroimaging phenotypes presented in Table 1 and a collection of 486 SNPs from 33 genes belonging to the top 40 Alzheimer's Disease (AD) candidate genes listed on the AlzGene database as of June 10, 2010. Section 6 concludes with a discussion of some ongoing work in this area.

\section{Mass Univariate and Voxel-Wise Approaches}

Mass univariate approaches avoid the complication of jointly modelling all neuroimaging phenotypes and genetic markers and simply conduct a test for association at each possible pair of voxel and genetic marker. Voxel-wise approaches are similar in that a separate model is fit independently at each voxel of the image, but these approaches may include multiple genetic markers in each model. The primary advantage of these approaches is that they make feasible a full brain-wide and genome-wide search for associations. 

We assume that neuroimaging and genetic data are available on $n$ subjects, where the imaging phenotype is denoted as $y_{\ell}(v)$, for the numerical value of the brain image of subject $\ell, \, \ell = 1, \dots, n$, at voxel $v, v = 1, \dots, V$. We denote the set of genetic markers for subject $l$ by $\mathbf{x}_{\ell} = (x_{\ell 1}, \dots , x_{\ell d})^{T}, \hspace{5pt}\ell = 1,\dots, n$, for a total of $d$ markers, where $x_{\ell j} \in \{0,1,2\}$ is the number of copies of the minor allele for the $j^{th}$ SNP, which takes values $x_{\ell j}=0$ (homozygotic major alleles), $x_{\ell j}=1$ (heterozygote), and $x_{\ell j}=2$ (homozygotic minor alleles). Finally, we let $\mathbf{z}_{\ell} = (z_{\ell 1}, \dots , z_{\ell p})^{T}, \hspace{5pt}\ell = 1,\dots, n,$ denote a collection of non-genetic variables for subject $l$.

Stein et al. (2010) is the first voxel-wise genome-wide association study (vGWAS) examining genetic influence on brain structure. The authors consider neuroimaging and genetic data obtained from $n = 818$ subjects as part of the ADNI. The neuroimaging data are based on brain MRI scans that are processed using an approach known as tensor-based morphometry (TBM). TBM (Ashburner et al., 2000) is used to create images representing volumetric tissue differences at each of approximately $31,000$ voxels for each individual, where the value of the image in a given voxel is obtained by calculating the determinant of the Jacobian matrix of a deformation that encodes local volume excess or deficit relative to an image that is representative of the sample known as the mean template image. The analysis relates the value of the image at each voxel to each of $448, 293$ SNPs. 

The statistical methodology considered by Stein et al. (2010) is fairly straightforward, though the resulting computation is still extensive due to the total number of tests considered. At each voxel $v$, a linear regression is conducted with response $y_{\ell}(v)$ (volumetric tissue difference relative to a mean template image at voxel $v$) and a separate regression model is fit relating this response to each SNP $x_{\ell j}$, assuming an additive genetic model. Additional independent variables age and gender are also included in the model, 
$$
y_{\ell}(v) = \beta_{0} + \beta_{1}\text{Age}_{\ell} + \beta_{2}\text{Gender}_{\ell} + \alpha x_{\ell j} + e_{\ell}(v, j).
$$ 
A standard p-value from this linear regression is obtained for each SNP-voxel pair (corresponding to a null hypothesis of $\alpha = 0$), and these p-values are computed at a given voxel as the model is fit repeatedly to all $d$ SNPs.  

To conserve memory, Stein et al. (2010) only save the minimum p-value at each voxel. Under the null hypothesis that the phenotype at a given voxel is not associated with any of the genetic markers, the minimum p-value computed at each voxel is not $\text{uniform}[0,1]$, but it is shown to be approximately $\text{Beta}(1, M_{eff})$, with $M_{eff} < M$, where $M_{eff}$ is the effective number of independent tests conducted at each voxel, and $M$ is the total number of genetic markers. The inequality $M_{eff} < M$ arises as a result of linkage disequilibrium. 

Stein et al. (2010) set the value of $M_{eff}$ equal to the number of principal components required to jointly explain 99.5\% of the variance in the SNPs. The  $\text{Beta}(1, M_{eff})$ distribution is used to correct the minimum p-value computed at each voxel via the probability integral transform so that the corrected minimum p-value is approximately distributed as $\text{uniform}[0,1]$. False Discovery Rate (FDR) procedures are then applied to adjust for multiple testing across voxels.  Under the proposed scheme the computations can be carried out in parallel across voxels, and Stein et al. (2010) employ 300 cluster nodes with a reported 27 hours of total computation time.

Hibar et al. (2011) develop a gene-based association method to complement single-marker GWAS for implicating underlying genetic variants in complex traits and diseases. The authors focus more broadly on gene-based approaches as they can be more powerful than traditional SNP-based approaches, with the relative power depending on how the genetic variants affect the phenotype. For example, if a gene contains multiple causal variants with small individual effects, SNP-based methods may miss these associations if a very stringent significance threshold is used. Gene-based tests also reduce the effective number of statistical tests by aggregating multiple SNP effects into a single test statistic. 

In Hibar et al. (2011) the SNP data are grouped by gene and SNPs that are not located in a gene are excluded. Considering a dataset of the same scale, both in terms of imaging and genetics, as that considered in Stein et al. (2010), after grouping SNPs a total of $18,044$ genes are left for analysis. The authors propose a gene-based association method that is based on principal components regression. Principal component analysis (PCA) is performed on the SNP data within each gene, storing all of the orthonormal basis vectors of the SNP matrix that explain a majority of the variance in the set of SNPs for a given gene. Basis vectors with the highest eigenvalues (higher proportions of explained variance) are included until 95\% of the SNP variability within the gene is explained, and the rest are discarded. The resulting `eigenSNPs' approximate the information in the collection of observed SNPs for a given gene. 

Hibar et al. (2011) apply their approach by examining associations between $18,044$ genes and approximately $31,000$ voxel-specific phenotypes. At each voxel, a multiple partial-F test is employed to test the the joint effect of all eigenSNPs for a given gene on the value of the image (volume difference) at the given voxel. The test is applied to all genes and the minimum p-value is recorded at each voxel. Inference then proceeds using an approximate Beta null distribution with FDR procedures applied to adjust for multiple testing as in Stein et al. (2010). The required computation across all voxels is parallelized over a cluster of 10 high performance 8-core CPU nodes and Hibar et al. (2011) report that the total time required to complete an analysis with their computational setup is approximately 13 days. Summarizing the SNP information in this way may have some disadvantages as well. In particular, if a single SNP has a large main effect, then testing the joint effect of all SNPs within that gene may dilute this association. However, when one considers the drastic reduction in the number of independent tests when comparing SNP-based linear regression with gene-specific summaries based on PCA, gene-based testing offers advantages when dealing with an extremely large number of voxels in an image phenotype.

In more recent work, Huang et al. (2015) have developed a fast voxel-wise genome-wide association analysis with an emphasis on large-scale imaging and genetic data. A key advantage of this methodology over the methods developed by Stein et al. (2010) and Hibar et al. (2011) is that it requires considerably less computational resources and is feasible to run on just a single CPU with reasonable processing time. The proposed approach is based on three main components: (1) a heteroscedastic linear model is specified at each voxel-locus pair; (2) a global sure independence screening procedure is incorporated to eliminate many noisy loci; (3) inference is based on wild bootstrap methods. The heteroscedastic linear model at voxel $v$ and locus $c$ takes the form
$$
y_{\ell}(v) = \mathbf{z}_{\ell} ^{T} \vbeta(v) + x_{\ell c} \alpha(c,v) + e_{\ell}(v), \,\, \ell = 1, \dots, n
$$
and the model makes no strong assumptions on $Var[e_{\ell}(v)]$, in particular, it may vary across subjects. The hypothesis test of interest is 
$$
H_{0}(c,v):  \alpha(c,v) = 0\,\, \text{versus} \,\, H_{1}(c,v):  \alpha(c,v) \ne 0 \,\, \text{for each} \,\, (c,v).
$$

Huang et al. (2015) introduce a standard Wald statistic $W(c,v)$ that is based on the ordinary least squares estimate of $\alpha(c,v)$. Under the heteroscedastic assumption of the regression model the standard approximations based on the $\chi^{2}_{1}$ (or $F$) distribution to the null distribution of $W(c,v)$ do not apply and a wild bootstrap method is proposed as an alternative. Huang et al. (2015) then focus on approximations that make such a procedure computationally feasible. 

A key aspect of these approximations is that a global sure independence screening procedure is used to eliminate many noisy loci. The global aspect of the screening procedure reduces the set of SNPs \emph{for all voxels simultaneously}. The authors define a global Wald statistic $W(c)$ for a given locus as the average of $W(c,v)$ taken over all voxels in the image. If, for a given locus $c$, it is the case that $H_{0}(c,v)$ holds for all voxels $v$, then Huang et al. (2015) argue that $W(c,v)$ asymptotically converges to a weighted $\chi^{2}$ distribution. The corresponding p-values are then computed for each locus $c$, and the top $N_{0}$ loci (e.g. $N_{0} = 1000$) are selected as the candidate set. 

Given the candidate set, a wild bootstrap approach is applied to determine significant voxel-locus pairs, or alternatively, significant cluster-locus pairs, where a cluster refers to a set of interconnected voxels each with test statistics exceeding a certain threshold. Allowing for cluster-wise inference in this way is an important advantage of this methodology over that proposed by Stein et al. (2010) and Hibar et al. (2011), as the voxel-specific tests proposed in the latter two articles might miss spatially extended signals that do not achieve significance at any isolated voxel. In this sense Huang et al. (2015) take advantage of the spatial information in the 3D images.

Ge et al. (2012) develop the first voxel-wise imaging genetics approach that allows for interactions between genetic markers. At each voxel the authors propose to fit a multi-locus model to associate the joint effect of several SNPs with the imaging trait at that voxel. The imaging traits are similar to those considered in Stein et al. (2010) though the model specified at each voxel is different. In particular, the semiparametric regression model specified at each voxel takes the form
$$
y_{\ell}(v) = \mathbf{z}_{\ell} ^{T} \vbeta(v) + h_{v}(\mathbf{x}_{\ell}) + e_{\ell}(v), \,\, \ell = 1, \dots, n,
$$
where $h_{v}(\mathbf{x}_{\ell})$ denotes a nonparametric function of the SNPs and the errors are assumed to be normally distributed with mean $0$ and standard deviation $\sigma_{v}$. In this case the non-genetic covariates (e.g., age, gender, education, handedness, and total intracranial volume) are modelled parametrically and the effect of genetic markers is modelled nonparametrically using a least squares kernel machines (Liu et al., 2007) approach. The function space containing $h_{v}(\cdot)$ is determined by an $n \times n$ kernel matrix which is a function of the genetic data and must be positive definite. The $(j, k)$ element of this matrix is a function of the SNP genotypes of subjects $j$ and $k$, and Ge et al. (2012) specify the form of this kernel to be
$$
k(\mathbf{x}_{j}, \mathbf{x}_{k}) = \frac{1}{2d}\sum_{s = 1}^{d}IBS(x_{j s}, x_{k s})
$$
where $IBS(x_{j s}, x_{k s})$ denotes the number of alleles shared identical by decent by subjects $j$ and $k$ at SNP $s$ and takes values $0$, $1$, or $2$. 

In this case the null hypothesis of interest is $H_{0}(v): h_{v}(\cdot) = 0$, which examines the effect of multiple SNPs at each voxel. Importantly, the model is very flexible and allows for interactions between the genetic markers. Ge et al. (2012) exploit a connection between least squares kernel machines and linear mixed models to derive a score statistic based on the null model (the model with no SNPs) and argue that this statistic follows a mixture of chi-squares under the null hypothesis. The score statistic has the advantage that its computation does not require the estimation of the function $h(\cdot)$. Using the Satterthwaite method, the distribution of this statistic under the null hypothesis is approximated by a scaled chi-squared distribution. 

Applying this technique at all voxels produces an image of score statistics and the authors assume that this statistic image behaves like a $\chi^{2}$ random field which facilitates inference using random field theory (Worsley, 1996). Random field theory (RFT) produces FWE-corrected p-values for voxel-wise and cluster-wise inference by accounting for the volume and smoothness of the statistic image. As RFT requires fairly strong assumptions on the statistic image and these assumptions may not be satisfied, the authors also develop voxel-wise inference based on permutation procedures with a parametric tail approximation based on the Generalized Pareto Distribution. 

Along with allowing for interactions among genetic variables the work of Ge et al. (2012) is the first to use RFT for inference in imaging genetics. Thus while correlation across voxels is not accounted for directly within a statistical model, the spatial structure of the imaging data is accounted for when computing FWE-corrected p-values using RFT.  

In a subsequent paper, Ge at al. (2015) extend the least squares kernel machine approach of Ge et al. (2012) to allow for both interactions between SNPs and further allow interactions between SNPs and non-genetic variables such as disease risk factors, environmental exposures, and epi-genetic markers. The model specified is of the form
$$
y_{\ell}(v) = \mathbf{z}_{\ell} ^{T} \vbeta(v) + h_{v, x}(\mathbf{x}_{\ell}) + h_{v, w}(\mathbf{w}_{\ell}) + h_{v, x, w}(\mathbf{x}_{\ell}, \mathbf{w}_{\ell}) + e_{\ell}(v), \,\, \ell = 1, \dots, n,
$$
where $ \mathbf{z}_{\ell}$ are non-genetic variables with linear effect and $\mathbf{w}_{\ell}$ are non-genetic variables with nonlinear effect that may interact with the genetic markers. As before a kernel machine based method is used to represent the nonparametric effects. In their application, Ge at al. (2015) only consider a scalar phenotype derived through MRI, namely, the hippocampal volume averaged between the two brain hemispheres; however, combining the voxel-wise inference of Ge et al. (2012) with the more flexible kernel machine model of Ge et al. (2015) seems feasible for dealing with phenotypes comprising an entire 3D image.

The mass univariate and voxel-wise approaches are appealing because of their simplicity and because the required univariate or multi-locus regression models are relatively easy to fit. Modelling the dependence between different voxels is avoided and this makes it feasible to perform large scale searches across many voxels of an image. Despite these advantages an important limitation is that these approaches do not exploit the spatial structure of phenotype-genotype associations. If a particular SNP is related to one voxel then it will likely be related to the neighbouring voxels as well, and these approaches do not allow us to borrow strength across voxels. This borrowing of strength can lead to higher power and is thus desired. Multivariate approaches are thus natural to consider, but these models typically require a significant reduction in the dimension of the neuroimaging phenotype by two orders of magnitude.

\section{Multivariate Approaches}
 
With multivariate approaches all of the neuroimaging phenotypes are included in a single large model that may account for the dependence structure across the different phenotypes while relating each of the phenotypes to all of the genetic markers. As a result, these approaches are typically not applied to imaging data at the voxel level as this is computationally intractable. A multivariate approach is typically applied to images reduced to a much coarser scale where each phenotype corresponds to a summary measure for an ROI in the brain. Table 1 provides an example of such summary measures for the 56 ROIs considered in our example. 

In the work of Wang et al. (2012) an estimator based on group sparse regularization is applied to multivariate regression for relating neuroimaging phenotypes to SNPs, where the SNPs are grouped by genes and this grouping structure is accounted for in the construction of the estimator. Let $\yl = (y_{\ell 1}, \dots , y_{ \ell c})^{T} \hspace{5pt}$ denote the imaging phenotype summarizing the structure of the brain over $c$ ROIs for subject $\ell$, $\ell = 1,\dots, n$. The corresponding genetic data are denoted by $\mathbf{x}_{\ell} = (x_{\ell 1}, \dots , x_{\ell d})^{T}, \hspace{5pt}\ell = 1,\dots, n$, where we have information on $d$ SNPs, and $x_{\ell j} \in \{0,1,2\}$ is the number of minor alleles for the $j^{th}$ SNP. We further assume that each SNP can be associated with a gene so that the set of genes represents a higher level grouping of the SNPs. Thus the set of SNPs can be partitioned into $K$ genes,  and we let  $\pi_k, k = 1,2, \dots, K$, denote the set containing the SNP indices corresponding to the $k^{th}$ group and $m_{k} = |\pi_{k}|$. This partitioning is used to allow for gene-wise association among SNPs. This is done through a regularization in which the coefficients of the SNPs within a gene, with respect to all of the imaging phenotypes, are penalized as a whole with an $l_{2}$-norm, while the  $l_{1}$-norm is used to sum up the gene-wise penalties to enforce sparsity between genes. The latter is important because in reality only a small fraction of genotypes are related to a specific phenotype.

It is assumed that $E(\mathbf{y_{\ell}}) = \mathbf{W}^{T}\mathbf{x}_\ell, \hspace{5pt} \ell = 1,\dots, n$, where $\mathbf{W}$ is a $d$ x $c$ matrix, with each row characterizing the association between a given SNP and the brain summary 
measures across all $c$ ROIs. The estimator proposed by Wang et al. (2012) takes the form
\begin{equation}
\label{Wang estimator} 
\hat{\mathbf{W}} = \underset{\mathbf{W}}\argmin   \mlikelihood  +\gamma_1 ||\mathbf{W}||_{G_{2,1}} +\gamma_2 ||\mathbf{W}||_{l_{2,1}}   
\end{equation}
where $\gamma_1$ and $\gamma_2$ are regularization parameters weighting a $G_{2,1}$-norm penalty \\ $||\mathbf{W}||_{G_{2,1}} = \sum_{k=1}^K  \sqrt{\sum_{i \in \pi_k}  \sum_{j=1}^c w_{ij}^2 }$ and an $\ell_{2,1}$-norm penalty $||\mathbf{W}||_{l_{2,1}} = \sum_{i=1}^{d}  \sqrt{ \sum_{j=1}^c w_{ij}^2}$ respectively. The $G_{2,1}$-norm encourages sparsity at the gene level. This regularization differs from group lasso (Yuan and Lin, 2006) as it penalizes regression coefficients for a group of SNPs across all imaging phenotypes jointly. As an important gene may contain irrelevant individual SNPs, or a less important
group may contain individually significant SNPs, the second penalty, an $\ell_{2,1}$-norm (Evgeniou and Pontil, 2007), is added to allow for additional structured sparsity at the level of SNPs (the rows of $\mathbf{W}$). 

The estimator (\ref{Wang estimator}) provides a novel approach for assessing associations between neuroimaging phenotypes and genetic variations as it accounts for several interrelated structures within genotyping and imaging data. Wang et al. (2012) develop an optimization algorithm for the computation of (\ref{Wang estimator}) and suggest the use of cross-validation for the selection of tuning parameters $\gamma_1$ and $\gamma_2$. A limitation of the proposed methodology is that it only furnishes a point estimate $\hat{\mathbf{W}}$ and techniques for obtaining valid standard errors or interval estimates are not provided.  
 
In recent work Greenlaw et al. (2017) address this limitation and extend the methodology of Wang et al. (2012) so that the uncertainty associated with $\hat{\mathbf{W}}$ can be quantified, and their methodology allows for formal statistical inference beyond the sparse point estimate $\hat{\mathbf{W}}$. Following the ideas of Park and Casella (2008) and Kyung et al. (2010), Greenlaw et al. (2017) develop a hierarchical Bayesian model that allows for full posterior inference. The Bayesian model is constructed with a particular prior for $\mathbf{W}$ so that the estimator (\ref{Wang estimator}) corresponds to the posterior mode. The spread of the posterior distribution then provides valid measures of posterior variability along with credible intervals for each regression parameter. 

Let $\mathbf{W}^{(k)} = (w_{ij})_{ i \in \pi_k}$ denote the $m_{k} \times c$ submatrix of $\mathbf{W}$ containing the rows corresponding to the $k^{th}$ gene, $k=1, \dots, K$. The hierarchical model of Greenlaw et al. (2017) corresponding to the estimator (\ref{Wang estimator}) takes the form 
\begin{equation}
\label{model - level 1}
\yl |\mathbf{W},\sigma^2  \distas{ind} MVN_c (\vW^{T} \vx_\ell \: , \: \sigma^2I_c),   \ell=1, \dots, n, 
\end{equation}
with the coefficients corresponding to different genes assumed conditionally independent
\begin{equation}
\label{model -level 2} 
\mathbf{W}^{(k)}| \lambda_{1}^{2}, \lambda_{2}^{2}, \sigma^2  \distas{ind} p(\mathbf{W}^{(k)}|  \lambda_{1}^{2}, \lambda_{2}^{2}, \sigma^2) \hspace{8pt} k=1,\dots,K, 
\end{equation}
and with the prior distribution for each $\mathbf{W}^{(k)}$ having a density function given by
\begin{equation}
\label{PML}
\begin{split}
p(\mathbf{W}^{(k)} |  \lambda_{1}^{2}, \lambda_{2}^{2}, \sigma^2) \propto \exp \left\lbrace - \frac{\lambda_{1}}{\sigma} \sqrt{ \sum_{i \in \pi_k} \sum_{j=1}^c w_{ij}^2 } \right\rbrace \\ \times \prod_{i \in \pi_k} \exp \left\lbrace -\frac{\lambda_{2}}{\sigma} \sqrt{\sum_{j=1}^c w_{ij}^2 } \right\rbrace.
\end{split}
\end{equation}
By construction, the posterior mode, conditional on $ \lambda_{1}^{2}, \lambda_{2}^{2}, \sigma^2$,  corresponding to the model hierarchy (2) - (4) is exactly the estimator (1) proposed by Wang et al. (2012) with $\gamma_{1} = 2\sigma \lambda_{1}$ and $\gamma_{2} = 2  \sigma \lambda_{2}$. This equivalence between the posterior mode and the estimator of Wang et al. (2012) is the motivation for the model; however, generalizations that allow for a more flexible covariance structure in (\ref{model - level 1}) can also be considered, and an extension of this model to allow for spatial correlation is discussed in Section 6. 

Greenlaw et al. (2017) develop a Gaussian scale mixture representation of this hierarchical model which allows for the implementation of Bayesian inference using a straightforward Gibbs sampling algorithm that is implemented in the R package `bgsmtr' (Bayesian Group Sparse Multi-Task Regression) which is available for download on CRAN (\url{https://cran.r-project.org/web/packages/bgsmtr/}). The selection of the tuning parameters $\lambda_{1}^{2}$ and $\lambda_{2}^{2}$ for this model is investigated in Nathoo et al. (2016), where selection of these tuning parameters based on a fully Bayes approach with hyperpriors, an empirical Bayes approach, and the WAIC are compared.
 
Vounou et al. (2010) propose an alternative strategy for multivariate regression modelling with imaging genetics data where the high-dimensional regression coefficient matrix is approximated by a low rank sparse matrix leading to a sparse reduced rank regression (sRRR) model. Suppose that $\mathbf{X}$ is the $n \times d$ design matrix of genetic markers and $\mathbf{Y}$ is the $n \times c$ matrix of imaging phenotypes. Beginning with the standard multivariate multiple linear regression model $\mathbf{Y} = \mathbf{X} \mathbf{C} + \mathbf{E}$, where $\mathbf{C}$ is the $d \times c$ matrix of regression coefficients, the approach proceeds by first imposing a rank condition on this matrix $\text{rank}(\mathbf{C}) \le \min(d,c)$ which leads to a decrease in the number of parameters that need to be estimated. In particular, if $\mathbf{C}$ has rank $r$ then it can be expressed as $\mathbf{C} = \mathbf{B} \mathbf{A}$ where $\mathbf{B}$ is $d \times r$ and $\mathbf{A}$ is $r \times c$ such that $\text{rank}(\mathbf{A}) = \text{rank}(\mathbf{B}) = r$. The loss function for estimation is based on the weighted least squares criterion
$
M = \text{Tr}\{(\mathbf{Y} - \mathbf{X}\mathbf{B} \mathbf{A})\Gamma(\mathbf{Y} - \mathbf{X}\mathbf{B} \mathbf{A})^{T}\},
$ 
where $\Gamma$ is a $c \times c$ positive definite weight matrix. Vounou et al. (2010) consider sparse estimation of both $\mathbf{B}$ and $\mathbf{A}$ through penalized estimation incorporating $l_{1}$-norm penalties. In particular, 
setting $\Gamma$ to be the identity matrix we have $$M = \text{Tr}\{\mathbf{Y}\mathbf{Y}^{T}\} - 2 \text{Tr}\{  \mathbf{A} \mathbf{Y}^{T} \mathbf{X}  \mathbf{B}\} + \text{Tr}\{  \mathbf{A} \mathbf{A}^{T} \mathbf{B}^{T} \mathbf{B} \},$$ where the first term on the RHS can be ignored as it does not depend on $\mathbf{B}$ or $\mathbf{A}$. Assuming $r=1$ and adding $l_{1}$-norm penalization yields the following optimization problem
$$
\argmin_{\mathbf{a}, \mathbf{b}} \{ -2    \mathbf{a} \mathbf{Y}^{T} \mathbf{X}  \mathbf{b}  +  \mathbf{a} \mathbf{a}^{T} \mathbf{b}^{T} \mathbf{b} + \lambda_{a}||\mathbf{a}^{T}||_{1}+ \lambda_{b}||\mathbf{b}^{T}||_{1} \}
$$
where  $\mathbf{a}$ is $1 \times c$ corresponding to the phenotypes and $\mathbf{b}$ is $d \times 1$ corresponding to the genetic markers. The sparsity of the solution depends on the values of $\lambda_{a}$ and $\lambda_{b}$ with the non-zero elements of $\hat{\mathbf{a}}$ selecting phenotypes and the non-zero elements of $\hat{\mathbf{b}}$ selecting genetic markers. 

The optimization problem is biconvex and Vounou et al. (2010) present an iterative algorithm for solving it. After the rank-one solution has been found, additional ranks can be obtained by applying the algorithm to the residuals of the data matrices. Vounou et al. (2010) suggest a graphical approach based on the residuals at each successive rank that can be used to select an optimal rank. As with the Wang et al. (2012) methodology, the methodology of Vounou et al. (2010) provides selection and point estimation but does not provide any mechanism for uncertainty quantification. This lack of uncertainty quantification can be a serious problem as we illustrate in our example application of Section 5.
 
Zhu et al. (2014) develop a Bayesian reduced rank model for imaging genetics that incorporates several enhancements above and beyond the methodology proposed in Vounou et al. (2010). First, the Bayesian approach enables uncertainty quantification for the regression parameters based on the posterior distribution. Second, in addition to reducing the dimension of the regression coefficient matrix with a low rank approximation, Zhu et al. (2014)  also incorporate a sparse latent factor model to represent the high-dimensional covariance matrix of the brain imaging phenotypes, with a multiplicative gamma process shrinkage prior assigned to the factor loadings. 

As with Vounou et al. (2010) the proposed model is based on the multivariate linear model $\mathbf{Y} = \mathbf{X} \mathbf{C} + \mathbf{E}$, where $\mathbf{E} = (\epsilon_{\ell k})$ and the rows of $\mathbf{E}$, each corresponding to a different subject, are assumed independent with $\vepsilon_{\ell} \sim MVN_{c}(\mathbf{0}, \Sigma)$. The rank $r$ decomposition of $\mathbf{C}$ with $r << \text{min}(c,d)$ takes the form $\mathbf{C} = \sum_{j=1}^{r} \mathbf{C}_{j}  = \sum_{j=1}^{r}\delta_{j} \mathbf{u}_{j}\mathbf{v}_{j}^{T}$, where $\mathbf{C}_{j} = \delta_{j} \mathbf{u}_{j}\mathbf{v}_{j}^{T}$ is the $j$-th layer, $\mathbf{u}_{j} \in \mathbb{R}^{d}$ and $\mathbf{v}_{j} \in \mathbb{R}^{c}$. The regression errors for each subject are expressed using a latent factor model $\vepsilon_{\ell} = \vLambda  \veta_{\ell} + \vxi_{\ell}$, where $\vLambda$ is a $d \times \infty$ factor loading matrix, $\veta_{\ell} \sim MVN_{\infty}(\mathbf{0},\mathbf{I}_{\infty})$, and 
$\vxi_{\ell} \sim MVN_{c}(\mathbf{0},\vSigma_{\xi})$ with $\vSigma_{\xi} = \text{diag}\{\sigma_{1}^{2},\dots,\sigma_{c}^{2}\}$. While it is typical to set the dimension of the latent factor $\veta_{\ell}$ to be much smaller than $\vepsilon_{\ell}$, the approach followed in  Zhu et al. (2014) is to choose a multiplicative gamma process prior for $\vLambda$ that shrinks increasingly the elements to zero as the column index increases, thereby avoiding the issue of choosing the number of factors (see also Bhattacharya and Dunson, 2011). The overall model for the imaging phenotype for a given subject can be written as
$$
\yl = \sum_{j=1}^{r}  X_{\ell}^{T} \delta_{j} \mathbf{u}_{j}\mathbf{v}_{j}^{T} + \vLambda  \veta_{\ell} + \vxi_{\ell},
$$
and Gaussian shrinkage priors are adopted for $\delta_{j}$, $\mathbf{u}_{j}$, and $\mathbf{v}_{j},\, j=1, \dots, r$. Zhu et al. (2014) present a Gibbs sampler that can be used to sample the posterior distribution and investigate a number of model selection criteria for choosing $r$. Their simulation studies indicate that the BIC outperforms several other model selection criteria in determining the true rank of $\mathbf{C}$.

Overall, the use of multivariate methods over the mass univariate and voxel-wise approaches can lead to greater efficiency through the borrowing of information across related brain imaging phenotypes. The approach of Wang et al. (2012) scales relatively well but does not provide uncertainty quantification. The Bayesian model of Greenlaw et al. (2017) addresses this issue at the expense of the greater computation required by the MCMC algorithm. As a result, the approach does not scale as well as that of Wang et al. (2012) and it requires parallel computation for the selection of tuning parameters. The sRRR approach proposed by Vounou et al. (2010) allows for potentially higher dimensional datasets with an appropriate choice of the rank of the regression coefficient matrix, while the Bayesian reduced rank approach of Zhu et al. (2014) offers several advantages including uncertainty quantification and a sparse latent factor model for the covariance matrix of the response. A disadvantage of the multivariate approaches, regardless of which is chosen, is that the imaging data must be substantially reduced to a summary measure over a reasonable number of ROIs (in the hundreds at most) while the mass univariate and voxel-wise approaches can be applied to tens of thousands of voxels. 

\section{Methods for Longitudinal Imaging Genetics Studies}

Longitudinal imaging genetics studies such as the ADNI study can provide insight into different rates of brain deterioration and how change in the structure of the brain over time is related to genetics. Szefer et al. (2017) have recently considered a longitudinal analysis of the ADNI database examining $75, 845$ SNPs in the Alzgene linkage regions and investigated associations with estimated rates of change in structural MRI measurements for 56 brain regions. Szefer et al. (2017) consider three phases of the ADNI study in their analysis, ADNI-1, ADNIGO, and ADNI-2. More information on the ADNI study including information on data access is available online at \url{http://adni.loni.usc.edu/about/}. The regions considered in Szefer et al. (2017) are the same as those considered in Greenlaw et al. (2017), and also described in Table 1 which we consider in our example analysis of the next session.

A primary innovation in the analysis of Szefer et al. (2017) is to construct from longitudinal MRI data and linear mixed models a set of subject and region specific rates of change over time. These estimated rates of change are then related to genetic markers using sparse canonical correlation analysis. Szefer et al. (2017) also use inverse probability weighting to account for the biased sampling design of the ADNI study, an aspect that has not been considered in many previous imaging genetics studies. 

Let $\yl(t) = (y_{\ell 1}(t), \dots , y_{ \ell c}(t))^{T} \hspace{5pt}$ denote the imaging phenotype summarizing the structure of the brain over $c$ ROIs for subject $\ell$, $\ell = 1,\dots, n,$ and at time $t$, where, for the ADNI study considered by Szefer et al. (2017) $t \in \{0,6,12,18,24\}$ months following entry into the study. For the $j^{th}$ ROI, Szefer et al. (2017) fit the following standard linear mixed model with random intercept and slope for time
\begin{multline}
y_{\ell j}(t) = \beta_{0j} + \beta_{1j}\text{MCI} + \beta_{2j}\text{AD} + \beta_{3j}t + \beta_{4j}\text{MCI}\times t \\+ \beta_{5j}\text{AD}\times t + \gamma_{1 \ell j} + \gamma_{2 \ell j} t + \epsilon_{\ell j}(t),
\end{multline}
where AD is an indicator for Alzheimer's Disease, MCI is an indicator for mild cognitive impairment, the $\beta$ terms denote fixed effects and the $\gamma$ terms denote random effects. The estimated rate of change extracted from the fitted linear mixed model is $\hat{\beta}_{3j} +   \hat{\beta}_{4j}\text{MCI} +  \hat{\beta}_{5j}\text{AD} + \hat{\gamma}_{2 \ell j}$, and these estimates, which are region specific, are used as the imaging phenotypes in the second stage  of their analysis after adjusting for population stratification using multidimensional scaling. The genetic markers are also adjusted for population stratification using the principal coordinates obtained from multidimensional scaling.

A sparse linear combination of the SNP genotypes that is most associated with a linear combination of the imaging phenotypes (the estimated rates of change) is obtained using sparse canonical correlation analysis (SCCA). SCCA is a multivariate method for estimating maximally correlated sparse linear combinations of the columns of two multivariate data sets. The degree of sparsity in the coefficients of the genotypes is controlled by a regularization parameter, and Szefer et al. (2017) choose this parameter so that approximately 10\% of the SNPs have non-zero coefficients. A bootstrap procedure is then used to estimate the relative importance of each SNP. In particular, sampling with replacement within each disease category (MCI, AD, Cognitively Normal), if $\vbeta_{b} = (\beta_{1b}, \dots, \beta_{d b})^{T}$ denotes the coefficient vector of the sparse linear combination of SNPs estimated from the $b^{th}$ bootstrap sample, Szefer et al. (2017) define the importance probability for the $k^{th}$ SNP as
$$
\text{VIP}_{k} = \frac{1}{B}\sum_{b=1}^{B}I\{\beta_{k b} = 0\}
$$
where $B$ is the total number of bootstrap samples. These importance probabilities are then used to select important subsets of SNPs. An interesting aspect of the analysis performed by Szefer et al. (2017) is that their procedure is applied to an ADNI-1 training sample to obtain subsets of important SNPs, and the authors are then able to validate many  of these priority SNPs using a validation set from ADNIGO/2.

An alternative model for longitudinal imaging genetics data has been proposed recently by Lu et al. (2017). The proposed model extends the Bayesian low rank model of Zhu et al. (2014) to the longitudinal setting. Unlike the two-stage longitudinal analysis of Szefer et al. (2017), the model of  Lu et al. (2017) links the time-varying neuroimaging data directly to the genetic markers in a single model that includes the data from all ROIs. Moreover, the proposed model allows for gene-age interactions so that the genetic effects on ROI volumes can vary across time. 

Letting $y_{\ell j}(t)$ denote the longitudinal imaging measure obtained from subject $\ell$ at ROI $j$ and time $t$, the model takes the form
$$
y_{\ell j}(t)  = \mathbf{X}_{\ell}^{T} \vbeta_{j} + \mu_{j}(t) + \mathbf{w}_{\ell}(t)^{T} \vgamma_{k} + \mathbf{z}_{\ell}(t)^{T} \vb_{\ell j} + \epsilon_{\ell j}(t), \,\, \ell = 1, \dots, n; j = 1, \dots c,
$$
where $ \mathbf{X}_{\ell}$ contains the genetic markers; $\mathbf{w}_{\ell}(t)$ is a vector of time-varying covariates that may include interactions between genetic markers and time;  $ \mu_{j}(t)$ is an overall temporal trend for the $j^{th}$ ROI; and $\vb_{\ell j}$ is a vector of subject specific Gaussian random effects for ROI $j$ corresponding to covariates $\mathbf{z}_{\ell}(t)$.  Lu et al. (2017) represent the functions $ \mu_{j}(t)$ using penalized-splines, and as in Zhu et al. (2014) a low rank approximation is used to approximate the regression coefficient matrix. The errors $\epsilon_{\ell j}(t)$ are represented through a sparse factor model 
$$
\epsilon_{\ell j}(t) = \vLambda  \veta_{\ell} (t)+ \vxi_{\ell}(t)
$$
with priors similar to those adopted in Zhu et al. (2014), including a multiplicative gamma process prior for $\vLambda$.  A Gibbs sampling algorithm is used to implement Bayesian inference. 

Overall, methods for longitudinal imaging genetics studies are just in their infancy, with very few published papers developing statistical methods to date. We believe there is significant scope for new work in this sub-area. Regarding the methods discussed here, a primary difference in the work of Lu et al. (2017) and that proposed by Szefer et al. (2017) is that the regression model on genetic markers in the latter case is built on estimated rates of change of the ROI volumes; whereas,  Lu et al. (2017) link the genetic data directly to the mean of the ROI volumes. Both methods provide useful and complimentary techniques for analyzing longitudinal imaging genetics data.

\section{Example Application}

We provide an example application examining an imaging genetics dataset obtained from the ADNI-1 database. The analysis presented here is considered in greater detail in Greenlaw et al. (2017); however, our objective in this case is simply to provide the reader with a simple example illustrating the use of some of the methods discussed in our review. 

The dataset includes both genetic and structural MRI data, the latter leading to the 56 imaging phenotypes presented in Table 1. The data are available for $n=632$ subjects (179 cognitively normal, 144 Alzheimer's, 309 mild cognitive impairment), and among all possible SNPs the analysis includes only those SNPs belonging to the top 40 Alzheimer's Disease (AD) candidate genes listed on the AlzGene database as of June 10, 2010. The data presented here are queried from the most recent genome build as of December 2014, from the ADNI-1 data. 

After quality control and imputation steps, the genetic data used for this study includes 486 SNPs from 33 genes and these genes along with the distribution of the number of SNPs within each gene is depicted in Figure 1. The freely available software package PLINK (Purcell et al., 2007) was used for genomic quality control.  Thresholds used for SNP and subject exclusion were the same as in Wang et al. (2012), with the following exceptions. For SNPs, we required a more conservative genotyping call rate of at least 95\% (Ge et al., 2012).  For subjects, we required at least one baseline and one follow-up MRI scan and excluded multivariate outliers. Sporadically missing genotypes at SNPs in the HapMap3 reference panel (Gibbs et al., 2003) were imputed into the data using IMPUTE2 (Howie et al., 2009).  Further details of the quality control and imputation procedure can be found in Szefer (2016). 

The MRI data from the ADNI-1 database are preprocessed using the FreeSurfer V4 software which conducts automated parcellation to define volumetric and cortical thickness values from the $c=56$ brain regions of interest that are detailed in Table 1. Each of the response variables are adjusted for age, gender, education, handedness, and baseline total intracranial volume (ICV) based on regression weights from healthy controls and are then scaled and centered to have zero-sample-mean and unit-sample-variance. 

We fit the Bayesian model of Greenlaw et al. (2017) and also compute the group sparse multi-task regression and feature selection estimator of Wang et al. (2012), both of which contain $56 \times 486 = 27,216$ regression parameters. For the former approach, we select potentially important SNPs by evaluating the 95\% equal-tail credible interval for each regression coefficient and select those SNPs where at least one of the associated credible intervals excludes 0. In total there are 45 SNPs and 152 regression coefficients for which this occurs. The 45 selected SNPs and the corresponding brain regions at which we see a potential association based on the 95\% credible intervals are listed in Table 2. 

Three SNPs, rs4311 from the ACE gene, rs405509 from the APOE gene, and rs10787010 from the SORCS1 gene stand out as being potentially associated with the largest number of ROIs. The 95\% credible intervals for the coefficients relating rs4311 to each of the $c=56$ imaging measures are depicted in Figure 2. Both the Bayesian posterior mean and the estimator of Wang et al. (2012) are indicated in the figure. 

In the original methodology of Wang et al. (2012) the authors suggest ranking and selecting SNPs by constructing a SNP weight based on the point estimate $\hat{\vW}$ and a sum of the absolute values of the estimated coefficients of each single SNP over all of the tasks. Doing so, the top 45 highest ranked SNPs contain 21 of the SNPs chosen using the Bayesian approach of Greenlaw et al. (2017) and these 21 SNPs are highlighted in Table 2 with bold font. The number 1 ranked (highest priority)  SNP using this approach is SNP rs3026841 from gene ECE1. In Figure 3 we display the corresponding point estimates for this SNP along with the 95\% credible intervals obtained from the Greenlaw et al. (2017) Bayesian approach, where again the credible intervals and point estimates are relating this SNP to each of the $c=56$ imaging measures. \emph{Importantly, we note that all 56 of the corresponding 95\% credible intervals include the value 0.} 

This result demonstrates the importance of accounting for posterior uncertainty beyond a sparse point estimate and illustrates the potential problems that may arise when estimation uncertainty is ignored, as in the approach of Wang et al. (2012). The methodology of Greenlaw et al. (2017) compliments the estimator of Wang et al. (2012) by providing uncertainty quantification, and both approaches may be applied together for such analyses.

While we have focussed on uncertainty quantification using credible intervals, the posterior distribution can be summarized through posterior probabilities of the form $Pr(|W_{ij}|> \delta|\text{Data})$ for known critical value $\delta >0$, or through kernel density estimation of the posterior density for certain regression coefficients. In the former case, adjustments for multiplicity can be made using Bayesian FDR procedures (Morris et al., 2008).

\section{DISCUSSION}

Imaging genetics is an emerging discipline that is based on combining two of the most important scientific fields where statistical research has made a major impact, genetics and neuroimaging. The resulting studies provide a number of big data challenges for statistical analysis. We have reviewed a variety of approaches for the analysis of such data focussing on mass univariate and voxel-wise approaches, multivariate approaches, and methods for longitudinal studies. Figure 5 summarizes these three approaches from a graphical perspective. 

One class of methods that we have not discussed in our review is the class of predictive methods for imaging genetics. In this setting the dependent variable is a condition or health outcome (such as presence of Alzheimer's disease), and the imaging and genetic data are used as features for classification. These methods typically aim at detecting prognostic markers. Typically, regularization methods, boosting algorithms and deep learning methods are applied to such problems (see e.g., Wang et al. 2012; Zhang et al. 2014). Within a Bayesian setting, a predictive model for imaging genetics with application to schizophrenia has been developed by Chekouo et al. (2016).  

While our review is not an exhaustive review of existing methods for imaging genetics, our aim was to provide the reader with a sample of the existing work and a flavour of the challenges for data analysis in this area. Indeed, this is a relatively new area in statistics and there is much scope for improving the existing methods. 

For example, one current avenue of interest is the extension of the methodology developed by Greenlaw et al. (2017) to accommodate a more realistic covariance structure for the imaging phenotypes. One approach for doing this is through a sparse latent factor model as considered in Zhu et al. (2014) and Lu et al. (2017). An alternative approach that we are currently investigating is the use of spatial models based on Markov random fields for the regression errors. More specifically, for the data considered in Greenlaw et al. (2017), our example in Section 5, and described in Table 1, the MRI-based phenotypes will exhibit two forms of correlation: (1) spatial correlation between neighbouring ROIs on the same brain hemisphere; and (2) correlation between corresponding measures on opposite brain hemispheres (e.g., the volume of the left hippocampus will be highly correlated with the volume of the right hippocampus). 

Considering the model formulation of Greenlaw et al. (2017), we begin by rearranging the imaging phenotypes so that they occur in left-right pairs in the vector $\yl \in \mathbb{R}^{c}$. Let $\mathbf{y}_{\ell, i} = (y_{\ell, i}^{(L)},y_{\ell, i}^{(R)})^{T}$ be the brain summary measures obtained at the $i^{th}$ ROI for both the left and right hemispheres. Then $\mathbf{y}_{\ell} = (\mathbf{y}_{\ell, 1}^{T}, \dots, \mathbf{y}_{\ell, \frac{c}{2}}^{T})^{T}$ is the imaging phenotype for subject $\ell$ with the elements rearranged so that left-right pairs are adjacent. The regression model is specified as $\mathbf{y}_{\ell} = \mathbf{W}^{T}\mathbf{x}_\ell + \ve_{\ell}$ where a spatial model for $\ve_{\ell}$ is based on a proper bivariate conditional autoregressive model (Gelfand and Vounatsou, 2013). 

We assume $\mathbf{A}$ is an adjacency matrix $A_{ij} \in \{0,1\}$ representing the spatial neighbourhood structure of ROIs on each hemisphere, with $\mathbf{D_{A}} = \text{diag}\{A_{i\cdot}, \, i=1,...,c/2\}$. The conditional specification for the regression errors is given by
$$
\ve_{\ell, i} | \{\ve_{\ell, j}, j \ne i\}, \rho, \Sigma \sim \text{BVN}(\frac{\rho}{A_{i\cdot}}\sum_{j=1}^{c/2}A_{ij}\ve_{\ell, j},\frac{1}{A_{i\cdot}}\Sigma ), \, i=1,\dots,c/2,
$$
where $\rho \in [0,1)$ characterizes spatial dependence and $\Sigma_{12}/\sqrt{\Sigma_{11}\Sigma_{22}} \in [-1,1]$ characterizes the dependence in the phenotypes across opposite hemispheres of the brain. 

The first level of the model can then be expressed as
$$ 
\mathbf{y}_{\ell} |\mathbf{W},\rho,\Sigma  \distas{ind} MVN_c (\mathbf{W}^{T}\mathbf{x}_\ell \: , \: (\mathbf{D_{A}} - \rho \mathbf{A})^{-1}\otimes \Sigma), \hspace{8pt} \ell=1, \dots, n 
$$
with higher levels of the model including the shrinkage prior for $\mathbf{W}$ specified as in Greenlaw et al. (2017) with some minor modifications. To clarify, the neighborhood structure and resulting adjacency matrix is used in the model to represent spatial dependence between phenotypes on the \emph{same} hemisphere of the brain, while $\Sigma$, a 2-by-2 matrix, is used to represent the correlation between the same phenotype on \emph{opposite} hemispheres of the brain. For specification of such a model it is convenient to arrange the phenotypes into left-right pairs which is why the rearrangement is needed. 

With regards to computation for this model, Greenlaw et al. (2017) and the corresponding R package 'bgsmtr' make use of sparse numerical linear algebra routines as the full conditional distributions for $\mathbf{W}$ have sparse precision matrices under that model. This is essential in order for the Gibbs sampling algorithm to be scalable to imaging genetics data of even moderately large size. In the proposed spatial model, so long as the adjacency matrix $\mathbf{A}$  is sparse the model structure still results in sparse precision matrices where required for faster computation. This is an advantage of using the Markov random field model over some other possible spatial models. The additional parameters $\rho \sim \text{Unif}(0,1)$ and $\Sigma \sim \text{inv-Wishart}(\mathbf{S},\nu)$ are easily added to the existing the Gibbs sampling algorithm. In addition to the use of Gibbs sampling, we are also developing a mean-field variational Bayes algorithm (see e.g., Nathoo et al., 2014) for the same model which should allow for greater scalability. We hope to report on results from this new model including a comparison of the different algorithms for Bayesian computation in a follow-up paper. The algorithms for fitting the new model will be available in the next version of the 'bgsmtr' R package.

\section*{ACKNOWLEDGEMENTS}
Linglong Kong and Farouk Nathoo are supported by funding from the Natural Sciences and Engineering Research Council of Canada and a CANSSI collaborative research team on neuroimaging data analysis. Farouk Nathoo holds a Tier II Canada Research Chair in Biostatistics for Spatial and High-Dimensional Data. 
Data collection and sharing for this project was funded by the Alzheimer's Disease Neuroimaging Initiative (ADNI) (National Institutes of Health Grant U01 AG024904) and DOD ADNI (Department of Defense award number W81XWH-12-2-0012). ADNI is funded by the National Institute on Aging, the National Institute of Biomedical Imaging and Bioengineering, and through generous contributions from the following: AbbVie, Alzheimer's Association; Alzheimer's Drug Discovery Foundation; Araclon Biotech; BioClinica, Inc.; Biogen; Bristol-Myers Squibb Company; CereSpir, Inc.; Cogstate; Eisai Inc.; Elan
Pharmaceuticals, Inc.; Eli Lilly and Company; EuroImmun; F. Hoffmann-La Roche Ltd and its affiliated company Genentech, Inc.; Fujirebio; GE Healthcare; IXICO Ltd.; Janssen Alzheimer Immunotherapy Research \& Development, LLC.; Johnson \& Johnson Pharmaceutical Research \& Development LLC.; Lumosity; Lundbeck; Merck \& Co., Inc.; Meso Scale Diagnostics, LLC.; NeuroRx Research; Neurotrack Technologies; Novartis Pharmaceuticals Corporation; Pfizer Inc.; Piramal Imaging; Servier; Takeda Pharmaceutical Company; and Transition Therapeutics. The Canadian Institutes of Health Research is providing funds to support ADNI clinical sites in Canada. Private sector contributions are facilitated by the Foundation for the National Institutes of Health (www.fnih.org). The grantee organization is the Northern California Institute for Research and Education, and the study is coordinated by the Alzheimer's Therapeutic Research Institute at the University of Southern California. ADNI data are disseminated by the Laboratory for Neuroimaging at the University of Southern California.

\newpage

\begin{longtable}[l]{lll}
\caption{
Imaging phenotypes defined as volumetric or cortical thickness measures of
$28 \times 2 = 56$ ROIs from automated Freesurfer parcellations. Each of the phenotypes in the table corresponds to two phenotypes in the data: one for the left hemisphere and the other for the right hemisphere.}\\ %\vspace*{5mm}
\hline
ID & Measurement & Region of interest\\
\hline
\endfirsthead
\hline
ID & Measurement & Region of interest\\
\hline
\endhead
\hline
\endfoot
AmygVol & Volume&Amygdala\\
CerebCtx & Volume&Cerebral cortex\\
CerebWM & Volume&Cerebral white matter\\
HippVol & Volume&Hippocampus\\
InfLatVent &Volume&Inferior lateral ventricle\\ 
LatVent &Volume&Lateral ventricle\\
EntCtx &Thickness&Entorhinal cortex\\
Fusiform &Thickness&Fusiform gyrus\\
InfParietal &Thickness&Inferior parietal gyrus\\
InfTemporal &Thickness&Inferior temporal gyrus\\ 
MidTemporal &Thickness&Middle temporal gyrus\\ 
Parahipp &Thickness&Parahippocampal gyrus\\
PostCing &Thickness&Posterior cingulate\\
Postcentral &Thickness&Postcentral gyrus\\ 
Precentral &Thickness&Precentral gyurs\\
Precuneus &Thickness&Precuneus\\
SupFrontal &Thickness&Superior frontal gyrus\\
SupParietal &Thickness&Superior parietal gyrus\\
SupTemporal &Thickness&Superior temporal gyrus\\
Supramarg &Thickness&Supramarginal gyrus\\
TemporalPole &Thickness&Temporal pole\\

MeanCing&Mean thickness&Caudal anterior cingulate, isthmus cingulate,\\ 
&&posterior cingulate, rostral anterior cingulate\\

MeanFront&Mean thickness& Caudal midfrontal\\ 
&&rostral midfrontal, superior frontal\\ 
&& lateral orbitofrontal, and medial orbitofrontal gyri\\ 
&&frontal pole\\

MeanLatTemp&Mean thickness&Inferior temporal, middle temporal\\
&&superior temporal gyri\\

MeanMedTemp&Mean thickness&Fusiform, parahippocampal, and lingual gyri,\\
&& temporal pole and transverse temporal pole\\

MeanPar&Mean thickness&Inferior and superior parietal gyri\\ 
&&supramarginal gyrus, and precuneus\\
MeanSensMotor&Mean thickness&Precentral and postcentral gyri\\ 
MeanTemp&Mean thickness&Inferior temporal, middle temporal, superior temporal,\\
&& fusiform, parahippocampal, lingual gyri, temporal pole,\\ 
&&transverse temporal pole\\
\hline
\end{longtable}%

\newpage

\begin{longtable}[l]{lll}
\caption{The 45 SNPs selected from the Bayesian model along with corresponding phenotypes where (L), (R), (L,R) denote that the phenotypes are on the left, right, and both hemispheres respectively. SNPs also ranked among the top 45 using the Wang et al. (2012) estimate are listed in bold.}\\ %\vspace*{5mm} 
\hline
SNP & Gene & Phenotype ID (Hemisphere)\\
\hline
\endfirsthead
\hline
SNP & Gene & Phenotype ID (Hemisphere)\\
\hline
\endhead
\hline
\endfoot
rs4305 &ACE& LatVent (R)\\

{\bf rs4311}&ACE&InfParietal (L,R)\\
&& MeanPar (L,R), Precuneus (L,R)\\
&& SupParietal (L), SupTemporal (L)\\
&& CerebCtx (R), MeanFront (R) \\ 
&&MeanSensMotor (R), MeanTemp (R)\\
&& Postcentral (R), PostCing (R)\\
&& Precentral (R), SupFrontal (R)\\ 
&&SupParietal (R)\\

{\bf rs405509} & APOE & AmygVol (L), CerebWM (L), Fusiform (L)\\
&& HippVol (L), InfParietal (L,R),SupFrontal (L,R), Supramarg (L,R) \\ 
&&InfTemporal (L), MeanFront (L,R), MeanLatTemp (L,R)\\
&& MeanMedTemp (L,R), MeanPar (L,R),\\
&&MeanSensMotor (L,R), MeanTemp (L,R)\\ 
&&MidTemporal (L,R), Postcentral (L,R), Precuneus (L,R)\\
&&SupTemporal (L,R), Precentral (R), SupParietal (R)\\

rs11191692 &	CALHM1&	EntCtx (L)\\

{\bf rs3811450} &	CHRNB2&	Precuneus (R)\\

rs9314349 &	CLU&	Parahipp (L)\\

{\bf rs2025935} &	CR1 & CerebWM (R), Fusiform (R), InfLatVent (R)\\

rs11141918 &	DAPK1&	CerebCtx (R) \\

{\bf rs1473180} &	DAPK1 & CerebCtx (L,R) ,EntCtx (L), Fusiform (L)\\ 
&& MeanMedTemp (L), MeanTemp (L), PostCing (L)\\

{\bf rs17399090} &	DAPK1 & MeanCing (R), PostCing (R)\\

rs3095747 &	DAPK1 &	InfLatVent (R)\\

{\bf rs3118846} &	DAPK1 &	InfParietal (R)\\

{\bf rs3124237} &	DAPK1 & PostCing (R), Precuneus (R), SupFrontal (R)\\

rs4878117	 & DAPK1 & MeanSensMotor (R), Postcentral (R)\\

rs212539	& ECE1 &	PostCing (R)\\

rs6584307 &	ENTPD7 & Parahipp (L)\\

rs11601726 &	GAB2 & CerebWM (L), LatVent (L)\\

{\bf rs16924159}&	IL33& MeanCing (L), PostCing (L), CerebWM (R)\\

rs928413	& IL33 &	InfLatVent (R)\\

{\bf rs1433099} &	LDLR& CerebCtx.adj (L), Precuneus (L,R)\\

rs2569537 &	LDLR& CerebWM (L,R)\\

rs12209631 &	NEDD9& CerebCtx (L), HippVol (L,R)\\

rs1475345 &	NEDD9 &	Parahipp (L)\\

{\bf rs17496723}&	NEDD9 &	Supramarg (L)\\

rs2327389 &	NEDD9 &	AmygVol (L) \\

{\bf rs744970} &	NEDD9 &	MeanFront (L), SupFrontal (L)\\

{\bf rs7938033} &	PICALM & EntCtx (R), HippVol (R)\\

{\bf rs2756271} &	PRNP & EntCtx (L), HippVol (L,R), InfTemporal (L), Parahipp (L)\\

{\bf rs6107516} &	PRNP & MidTemporal (L,R)\\

rs1023024 &	SORCS1 & MeanSensMotor (L), Precentral (L)\\

{\bf rs10787010} &	SORCS1 & AmygVol (L), EntCtx (L,R)\\ 
&&MeanFront (L), Fusiform (L)\\
&&HippVol (L,R), InfLatVent (L), InfTemporal (L)\\ 
&&MeanMedTemp (L,R), MeanTemp (L)\\ 
&&Precentral (L), TemporalPole (R)\\

rs10787011 &	SORCS1& EntCtx (L,R), HippVol(R)\\

rs12248379 &	SORCS1	& PostCing (R)\\

rs1269918 &	SORCS1 &CerebCtx (L), CerebWM (L), InfLatVent (L)\\

{\bf rs1556758} &	SORCS1	&SupParietal (L) \\

{\bf rs2149196} &	SORCS1 & MeanSensMotor (L), Postcentral (L,R)\\

{\bf rs2418811} &	SORCS1 & CerebWM (L,R), InfLatVent.adj (L)\\

{\bf rs10502262} &	SORL1 & MeanCing (L), InfTemporal (R), Supramarg (R)\\

{\bf rs1699102} &	SORL1 & MeanMedTemp (R),  MeanTemp (R)\\

rs1699105 &	SORL1 & MeanCing (L), Precuneus (L)\\

rs4935774&	SORL1	&CerebWM (L,R)\\

rs666004&	SORL1&	InfTemporal (L)\\

rs1568400&	THRA& Precentral (L), TemporalPole (R)\\

rs3744805&	THRA& MeanSensMotor (R), Postcentral (R), Precentral (R)\\

rs7219773&	TNK1& MeanSensMotor (L), Precentral (L), Postcentral (R)\\
\hline
\end{longtable}

\newpage

\begin{figure}[htbp]
\centering
\includegraphics[width=\textwidth]{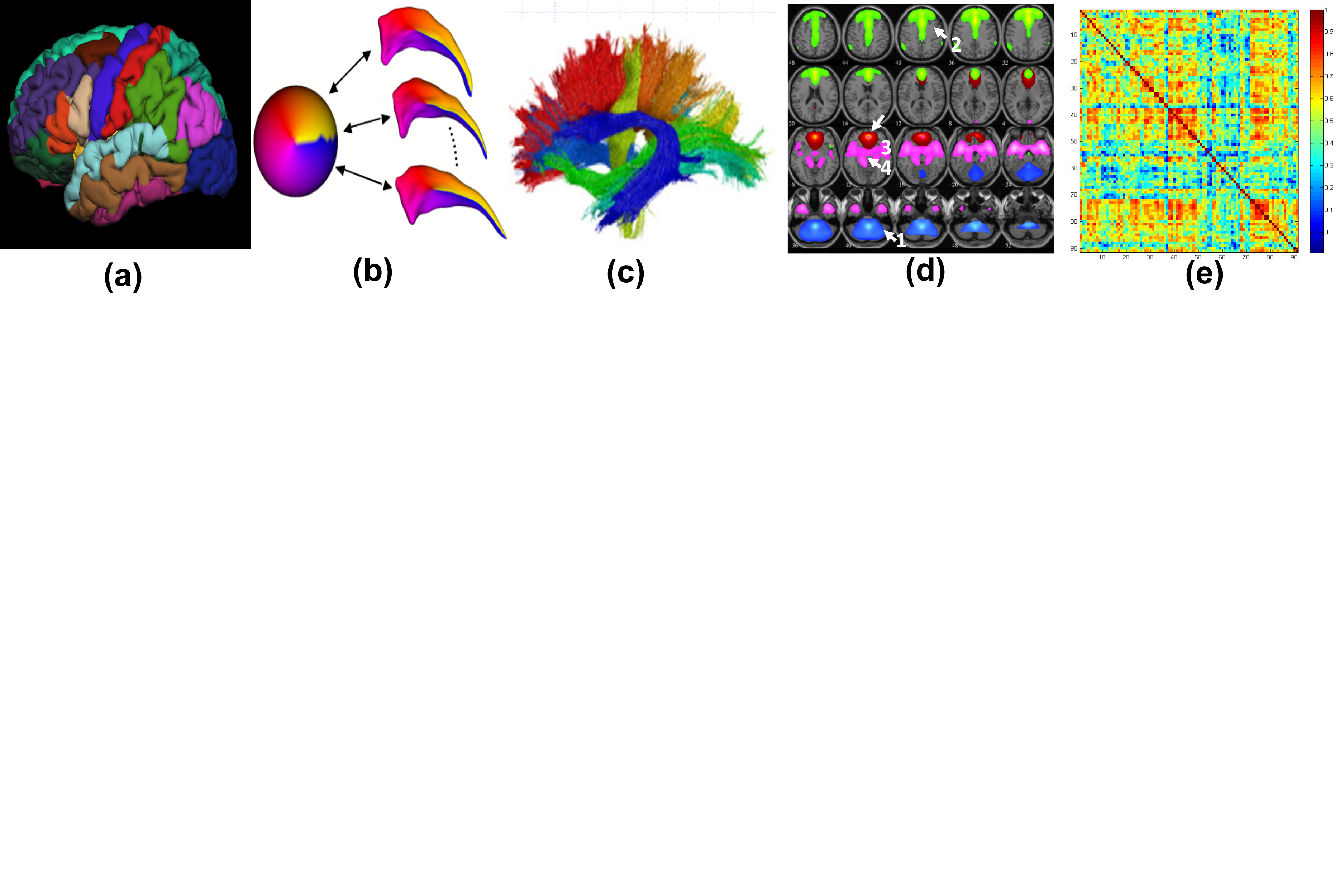}
\caption{Selected  Imaging Phenotypes (IPs)}
\end{figure}

\begin{figure}[htbp]
\centering
\includegraphics[scale=0.45]{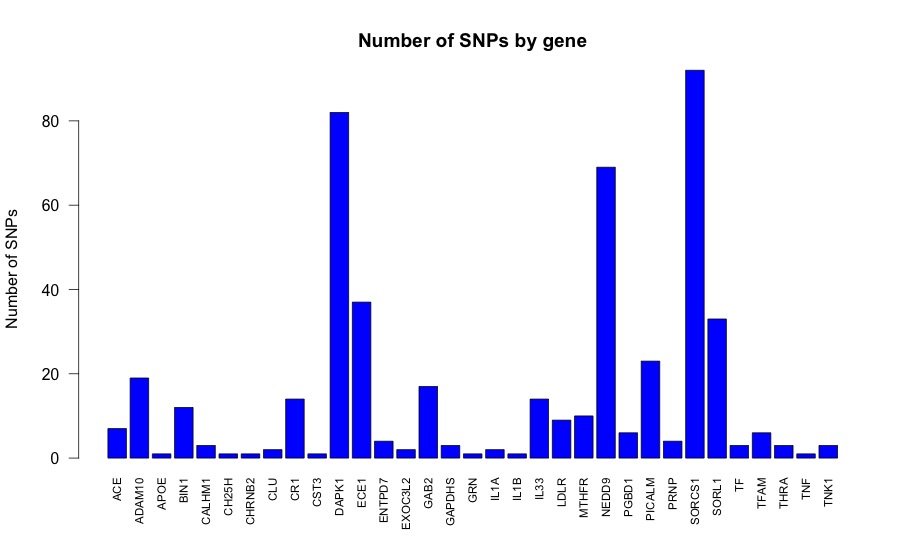}
\caption{Each of the 33 genes partitioning the 486 SNPs included in the example data analysis of Section 5.}
\end{figure}

\begin{figure}[htbp]
\centering
\includegraphics[scale=0.55]{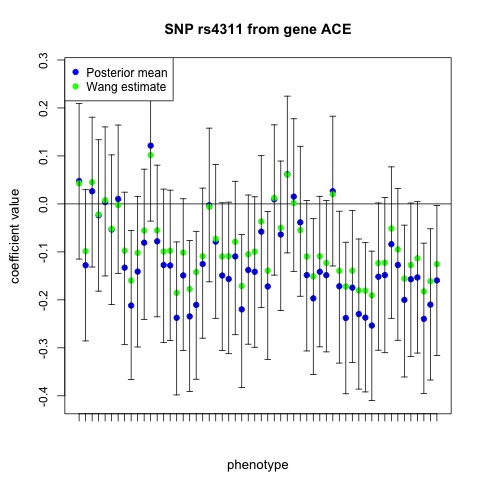}
\caption{The 95\% equal-tail credible intervals relating the SNP rs4311 from ACE to each of the $c=56$ imaging phenotypes. Each imaging phenotype is represented on the x-axis with a tick mark and these are ordered in the same order as the phenotypes are listed in the rows of Table 1, first for the left hemisphere and then followed by the same phenotypes for the right hemisphere.}
\end{figure}

\begin{figure}[htbp]
\centering
\includegraphics[scale=0.55]{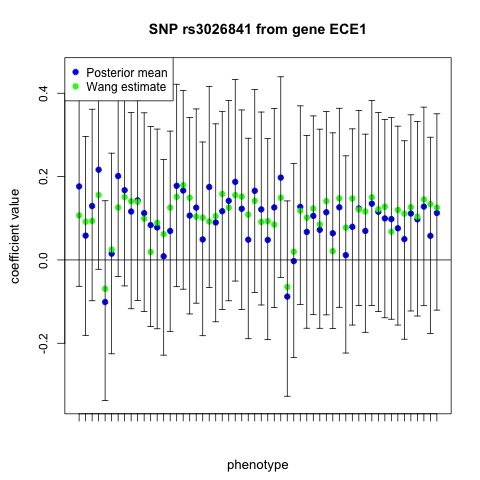}
\caption{The 95\% equal-tail credible intervals relating the SNP rs3026841 from ECE1 to each of the $c=56$ imaging phenotypes. Each imaging phenotype is represented on the x-axis with a tick mark and these are ordered in the same order as the phenotypes are listed in the rows of Table 1, first for the left hemisphere and then followed by the same phenotypes for the right hemisphere.}
\end{figure}

\begin{figure}[htbp]
\centering
\includegraphics[width=\textwidth]{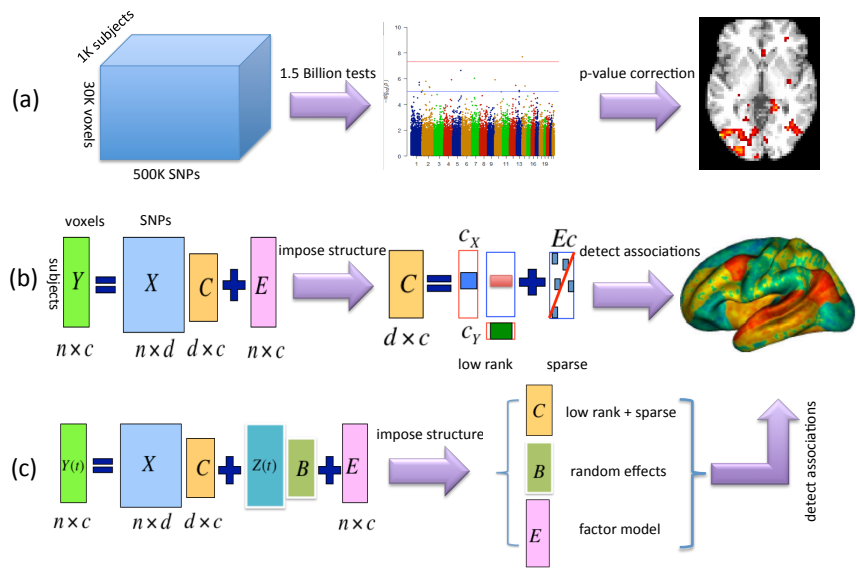}
\caption{The three approaches discussed in the paper summarized from a graphical perspective: (a) mass univariate and voxel-wise approaches; (b) multivariate approaches; (c) methods for longitudinal imaging genetics studies.}
\end{figure}

\end{document}